\documentclass{article}[12pt]

\title{A Component-wise EM Algorithm for Mixtures}
\date{}
\usepackage{amssymb,amsmath}
\usepackage{epsfig}
\usepackage{epsf}

\newcommand{\Sum}[2]{{\displaystyle \sum^{#2}_{#1}}}

\newtheorem{lemma}{Lemma}

\newtheorem{proposition}{Proposition}
\newtheorem{assumption}{Assumption}
\newtheorem{theorem}{Theorem}
\newtheorem{corollary}{Corollary}
\begin{document}
\maketitle

{\bf Abstract}: Estimation
finite mixture distributions is typically an
incomplete data structure problem for which the  Expectation-Maximization
(EM) algorithm
can be used. One drawback of the algorithm is its slow convergence 
in some situations. In the mixtures case, little progress in 
speeding up EM has been made. 
Standard EM procedures update all parameters simultaneously. In the missing
data context, it has been shown that sequential updating could lead to
faster convergence. 
In this paper we 
 propose a component-wise EM for mixtures, which updates the parameters sequentially. 
It intrinsically decouples the parameter updates so that
 the estimated proportions may not sum to 1  during an iteration. 
While maintaining monotone convergence, the algorithm may leave the 
parameter space but is guaranteed to return upon convergence. 
We give an interpretation of this procedure as  a proximal point algorithm 
and use it to prove the convergence.
Illustrative numerical experiments show how our algorithm compares to EM and
a version of the SAGE algorithm. [un mot sur les perf]
\newpage
\section{Introduction}
Estimation in finite mixture distributions is typically an
incomplete data structure problem for which the EM algorithm 
is used (see for instance Dempster, Laird and Rubin 1977 and
 Redner and Walker 1984).
The most documented problem occuring with the EM algorithm is its 
possible low speed in some situations.
Many 
papers have proposed extensions of
the EM algorithm based on standard numerical tools to speed up the
convergence. Possible references include
 Louis (1982), Lewitt and Muehllehner (1986), Kaufman (1987),
Meilijson (1989), and  Jamshidian and Jennrich (1993). 
There are often effective, but they do not guarantee monotone
increase in the objective function. To overcome this problem,  alternatives
 based on model reduction and efficient data augmentation 
 have recently been considered. As regards model reduction, 
we  refer to 
 Meng and Rubin (1993), Liu and Rubin (1994).
For 
data augmentation, see  Fessler and Hero (1994, 1995), Hero and Fessler 
(1995),
Meng and van Dyk (1997, 1998), Neal and Hinton (1998), Liu, Rubin and Wu 
(1998),
see also the chapter 5 of McLachlan (1997). These 
extensions share the simplicity and stability with EM while speeding up the
convergence. However, as far as we know, 
only two extensions were devoted to 
speeding up the convergence in the mixture case which is one of
the most important domains of application for EM 
(Pilla and Lindsay 1996, Liu and Sun 1997). The first one of 
Pilla and Lindsay (1996) is based on a restricted efficient data augmentation 
scheme for the estimation of the proportions for known discrete distributions.
While the second extension of Liu and Sun (1997) is concerned with the 
implementation of the ECME algorithm (Liu and Rubin 1994) for mixture distributions.

In this paper we propose, study and illustrate
a component-wise EM algorithm (CEMM: Component-wise
 EM algorithm for Mixtures)
aiming at overcoming the slow convergence problem
in the finite mixture context.  
Our approach is based on a recent work of Chr\'etien and Hero (1998a,b, 1999), which recasts the EM procedure in the framework
of proximal point algorithms Rockafellar (1976a) and Teboulle (1997).
In Section \ref{emsec} we present the EM algorithm
for mixtures.
In Section \ref{cem2sec},
 we describe our component-wise algorithm and show
that it can be interpreted as a proximal point algorithm. 
Using this interpretation, convergence of CEMM is proved in
the same section. 
Illustrative numerical experiments comparing the
behaviors of EM, a version of the  SAGE algorithm 
(Fessler and Hero 1994, Fessler and Hero 1995)
 and CEMM are presented in Section \ref{numsec}. Concluding remarks
end the paper. An appendix carefully describes
the SAGE method in the mixture context in order to provide
detailed comparison with the proposed CEMM.
\section{The EM algorithm for mixtures}
\label{emsec}

We consider a $J$-component mixture in ${\mathbb R}^{d}$
\begin{equation} \label{mix}
g(y | \theta) = \sum_{j = 1}^{J} p_{j} \varphi (y | \alpha_{j})
\end{equation}
where  the $p_j$'s ($0 < p_{j} < 1$ and $\Sum{j=1}{J} 
p_{j} = 1$) are the mixing proportions and  where $\varphi( y
| \alpha)$ is a density function parametrized by $\alpha$. 
The vector parameter to be
estimated is $\theta = (p_{1}, \ldots, p_{J}, \alpha_{1}, \ldots,
\alpha_{J}).$ 
The parametric families of mixture densities are assumed to be identifiable. 
This means that  
 for any two members of the form ({\ref{mix}), 
\begin{equation*}
g(y|\theta)\equiv g(y |\theta')
\end{equation*}
if and only if $J=J'$ and we can permute the components labels so that
$p_j=p_{j'}$ and $\varphi(y|\alpha_j)=\varphi(y|\alpha_{j'})$,
 for $j=1,\ldots,J$. 
Most mixtures of interest are identifiable (see for instance Redner and Walker 1984).
For the sake of simplicity, we restrict the present analysis to Gaussian mixtures,
but extension to more general mixtures is straightforward. 
Thus, $\varphi(y | \mu, \Sigma)$ denotes the density of a Gaussian distribution with
mean $\mu$ and variance matrix $\Sigma$. The parameter to be
estimated is \begin{equation*}\theta = (p_{1}, \ldots, p_{J}, \mu_{1}, \ldots,
\mu_{J}, \Sigma_{1},
\ldots,\Sigma_{J}).\end{equation*}
In the following, we denote $\theta_j=(p_j,\mu_j,\Sigma_j)$,
 for $j=1,\ldots,J$. We also denote by $\Theta$ the parameter space
$\{(p_1,\ldots,p_J,\mu_1,\ldots,\mu_J,\Sigma_1,\ldots,\Sigma_J)\}$
and by $\Theta^\prime$ the affine submanifold
\begin{equation} 
\Theta^\prime=\Big\{\theta\in \Theta \mid \sum_{\ell=1}^Jp_\ell=1 \Big\}. \nonumber
\end{equation}

The mixture density estimation problem is typically a missing data problem
for which the EM algorithm appears to be useful. 

Let ${\mathbf y}= (y_{1}, \ldots, y_{n}) \in {\mathbb R}^{dn}$ be
an observed sample from the mixture distribution $g(y | \theta)$.
We assume that the component from which each $y_{i}$ arised  is
unknown so that the missing data are the labels $z_i$, $i=1, \ldots, n$.
We have $z_{i}=j$
if and only if
$j$ is the mixture component from which $y_{i}$ arises.
Let ${\mathbf z}= (z_1,\ldots,z_n)$ denote the missing
data, ${\mathbf z} \in B^n$, where $B=\{1,\ldots,J\}$. 
The complete sample is ${\mathbf x} =
(x_{1}, \ldots, x_{n})$ with $x_{i} = (y_{i}, z_{i})$ and we have
${\mathbf x}=({\mathbf y}, {\mathbf z})$. 
The observed log-likelihood is
\begin{equation*}
L(\theta | {\mathbf y})= \log{\mathbf g}({\mathbf y} | \theta),
\end{equation*}
where ${\mathbf g}({\mathbf y} | \theta)$ denotes the density of
the observed sample ${\mathbf y}$. Using (\ref{mix}) leads to  
\begin{equation*} L(\theta | {\mathbf y}) = \sum_{i=1}^{n} \log \left\{ \sum_{j =
1}^{J} p_{j} \varphi (y_{i} | \mu_{j}, \Sigma_{j})\right\}.\end{equation*}
The complete log-likelihood is
\begin{equation*}
L(\theta | {\mathbf x})= \log{\mathbf f}({\mathbf x} | \theta),
\end{equation*}
where ${\mathbf f}({\mathbf x} | \theta)$ denotes the density of
the complete sample ${\mathbf x}$. We have
\begin{equation} \label{loglik}
 L(\theta | {\mathbf x}) = \sum_{i=1}^{n} \left\{ \log p_{z_i} + \log
\varphi (y_{i} | \mu_{z_i}, \Sigma_{z_i})\right\}.\end{equation}
The conditional density function of the complete data given ${\mathbf y}$
\begin{equation} \label{tc}
{\mathbf t}({\mathbf x} | {\mathbf y}, \theta)= \frac
{{\mathbf f}({\mathbf x} | \theta)}{{\mathbf g}({\mathbf y} | \theta)}\end{equation}
takes the form
\begin{equation} \label{tfacto}
 {\mathbf t} ({\mathbf x} | {\mathbf y}, \theta) = \prod_{i=1}^{n} t_{iz_{i}}
 (\theta) \end{equation}
where $t_{ij} (\theta), j = 1, \ldots, J$ denotes the
conditional probability, given ${\mathbf y}$, that $y_{i}$ arises
from the mixture component with density $\varphi (. |\mu_{j}, \Sigma_{j})$. 
>From Bayes formula, we have for each
$i$ $(1 \leq i \leq n)$ and $j$ $(1 \leq j \leq J)$
\begin{equation}\label{tij}
t_{ij} (\theta) = {p_{j} \varphi (y_{i} | \mu_{j},
\Sigma_{j}) \over {\displaystyle \sum_{\ell=1}^{J}} p_{\ell} \varphi (y_{i} |
\mu_{\ell}, \Sigma _{\ell})}\quad . \end{equation}
Thus the conditional expectation of the complete log-likelihood given
${\mathbf y}$ and a previous estimate of $\theta$, denoted $\theta'$,
\begin{equation*}
Q(\theta | \theta')= I\!\!E\left[\log {\mathbf f}(\theta | {\mathbf x})|{\mathbf y},
 \theta'\right]
\end{equation*}
takes the form
\begin{equation} \label{qval}
 Q(\theta|\theta') = \sum_{i=1}^{n} \sum_{\ell = 1}^{J} t_{i\ell} (\theta') 
\left\{\log p_{\ell} + \log \varphi (y_{i} | \mu_{\ell},
\Sigma_{\ell}) \right\}.
\end{equation}
The EM algorithm generates a sequence of approximations to derive the
maximum observed likelihood estimator starting from an initial
guess $\theta^0$,  using two steps. The  $k$th iteration is as follows\\
\mbox{}\\
{\bf E-step}: Compute $Q(\theta | \theta^k)=
I\!\!E\left[\log {\mathbf f}({\mathbf x}|\theta)|{\mathbf y}, \theta^k\right]$.\\
{\bf M-step}: Find $\theta^{k+1}= \arg\max\limits_{\theta \in \Theta^\prime} Q(\theta | \theta^k)$,\\
\mbox{}\\
 In many situations, including the mixture case, the explicit
 computation of $Q(\theta | \theta^k)$
in the E-step is unnecessary and this step reduces to the computation
of the conditional density 
${\mathbf t}({\mathbf x}| {\mathbf y},\theta^k)$.
For Gaussian mixtures, these two steps take the form\\
\mbox{}\\
{\bf E-step}: For $i = 1,\ldots, n$ and $j= 1, \ldots, J$
compute
\begin{equation} \label{eem}
t_{ij} (\theta^{k}) = {p_{j}^{k}
\varphi (y_{i} | \mu_{j}^{k}, \Sigma_{j}^{k}) \over
{\displaystyle \sum_{\ell=1}^{J}} p_{\ell}^{k} \varphi (y_{i} | \mu_{\ell}^{k},
\Sigma_{\ell}^{k})}.
\end{equation}
{\bf M-step} : Set  $\theta^{k+1} =(p_1^{k+1},\ldots,p_J^{k+1},
\mu_1^{k+1},\ldots,\mu_J^{k+1},\Sigma_1^{k+1},\ldots,\Sigma_J^{k+1})$
with 
\begin{equation} \label{mem}
\begin{array}{l}
\displaystyle{p_j^{k+1}=\frac1n \sum_{i=1}^n t_{ij}(\theta^k)}\\
\\
\displaystyle{\mu_j^{k+1}=\frac{\Sum{i=1}{n} t_{ij}(\theta^k) \;y_i}
{\Sum{i=1}{n} t_{ij}(\theta^k)}}\\
\\
\displaystyle{\Sigma_j^{k+1}=\frac{\Sum{i=1}{n} t_{ij}(\theta^k)
(y_i-\mu_j^{k+1})(y_i-\mu_j^{k+1})^{\sf T}}{\Sum{i=1}{n} t_{ij}(\theta^k)}}.
\end{array}
\end{equation}
Note that at each iteration, the following properties hold
\begin{equation*}
\mbox{for } i=1,\ldots,n,  \quad \sum_{j=1}^J t_{ij}(\theta^k)=1
\end{equation*}
\begin{equation} \label{cons}
\mbox{and } \sum_{j=1}^Jp_j^k=1.
\end{equation}

\section{A Component-wise EM for mixtures}
\label{cem2sec}

Serial decomposition of optimization methods is a well known procedure in
 numerical analysis.
Assuming that $\theta$ lies in ${\mathbb R^p}$,
the optimization problem
\begin{equation*}
\max\limits_{\theta\in \mathbb R^p} \Phi(\theta)
\end{equation*}
is decomposed into a series of coordinate-wise maximization problems of the
form
\begin{equation*}
\max\limits_{\eta \in \mathbb R} \Phi(\theta_1,\ldots,\theta_{j-1},\eta,\theta
_{j+1},\ldots,
\theta_p).
\end{equation*}
This procedure is called a Gauss-Seidel scheme. The study of this method is 
standard (see Ciarlet 1988 for example). 
It has the advantage of using the new information
as soon as it is available rather than waiting until all parameters have 
been updated.
One of the most promising general purpose extension of EM, going in this
direction, is the
 Space-Alternating Generalized EM
(SAGE) algorithm of Fessler and Hero (1994).
Improved convergence rates are reached by updating the parameters
sequentially in small groups associated
to small missing data spaces rather than one large
complete data space.
The idea is that less informative missing data spaces lead to smaller
root-convergence factors and hence faster converging algorithms.
General description and details concerning the
rationale, the properties and illustrations of the SAGE algorithm
can be found in Fessler and Hero (1994,1995), Hero and Fessler (1995).
The CEMM algorithm  is closely related to the SAGE approach. 
For comparison purpose, we described in the appendix of
 Celeux and al. (1999),  
a version  of SAGE for Gaussian mixtures.
This version is nearly  a component-wise algorithm
except that the mixing proportions need to be
 updated in the same iteration, which
involves the whole complete data structure.
For this reason, it may not be significantly  faster than
 the standard EM algorithm.
This points out the main interest of
the component-wise EM algorithm that we propose  for mixtures.
No iteration needs the whole complete data space
as missing data space. It can
therefore be expected to converge faster in various situations.
 
\subsection{The CEMM algorithm}
\label{cem2algo}

Our Component-wise EM algorithm for Mixtures (CEMM)
considers the decomposition of the parameter vector
$\theta=(\theta_j,j=1,...,J)$ with $\theta_j=(p_j,\mu_j,\Sigma_j)$.
The idea is to update only one component at a time, letting the other
parameters unchanged. The order according to which the components are
visited may be arbitrary, prescribed or varying adaptively.
 For simplicity, in our presentation,
 the components are  updated successively,
starting from $j=1, \ldots, J$ and repeating this after $J$ iterations.
Therefore the component updated at iteration $k$ is given by (\ref{defj})
 and the $k$th iteration of the  algorithm is as follows.
For
\begin{equation}\label{defj}
j=k-\frac{k}J\rfloor J+1,
\end{equation}
$.\rfloor $ denoting the integer part, it alternates the following steps
 
\noindent
{\bf E-step}:
Compute for $i=1, \ldots, n$,
\begin{equation} \label{ecem}
t_{ij} (\theta^{k}) = {p_{j}^{k}
\varphi (y_{i} | \mu_{j}^{k}, \Sigma_{j}^{k}) \over
{\displaystyle \sum_{\ell=1}^{J}} p_{\ell}^{k} \varphi (y_{i} | \mu_{\ell}^{k}
,
\Sigma_{\ell}^{k})}.
\end{equation}
 
\noindent
{\bf M-step}: Set
\begin{equation} \label{mcem}
\begin{array}{l}
\displaystyle{p_{j}^{k+1}=\frac1n \sum_{i=1}^n t_{ij}(\theta^k)}\\
\\
\displaystyle{\mu_{j}^{k+1}=\frac{\Sum{i=1}{n} t_{ij}(\theta^k)y_i}{\Sum{i=1}{
n} t_{ij}(\theta^k)}}\\
\\
\displaystyle{\Sigma_{j}^{k+1}=\frac{\Sum{i=1}{n} t_{ij}(\theta^k)
(y_i-\mu_j^{k+1})(y_i-\mu_j^{k+1})^{\sf T}}{\Sum{i=1}{n} t_{ij}(\theta^k)},}
\end{array}
\end{equation}
and for $\ell \neq j$,
$\theta^{k+1}_\ell=\theta^k_\ell\;.$
 
\vspace{.3cm}
 
Note that the main difference with the SAGE algorithm presented in 
Celeux and al. (1999) is that the updating steps of the mixing
proportions cannot be regarded as maximization steps as in SAGE.
Also, in CEMM, the $p_j$'s in (\ref{mcem}) do not necessarily sum to 1, so
that the algorithm may leave the parameter space. 
Consequently, the SAGE
standard assumptions are not
satisfied and a specific convergence analysis must be
achieved. It shows that the CEMM algorithm is guaranteed to return in the
parameter space upon convergence.
  It is based on the proximal interpretation of CEMM
 given in the
next subsection.

\subsection{Lagrangian and Proximal representation of CEMM}
\label{proxsec}

As shown in Chr\'etien  and Hero (1998a),
the EM procedure can be recast into a proximal point framework.
 This point of view  provides much insight into the algorithm  convergence
 properties (see Chr\'etien and Hero 1999).
The proximal point algorithm was first studied in
Rockafellar (1976b). The proximal methodology was then applied to many
types of algorithms and is still in great effervescence
(see Teboulle 1992,1997 for instance and the literature
therein).
Considering
 the general problem of maximizing a concave function $\Phi(\theta)$
on ${\mathbb R}^p$,
the proximal point algorithm is an iterative procedure which is defined
by the following recurrence,
\begin{equation}
\label{proxit}
\theta^{k+1}=\arg\max\limits_{\theta \in {\mathbb R}^p}\left\{\Phi(\theta)
-\frac{1}{2} \|\theta-\theta^k\|^2\right\}.
\end{equation}
In other words, the objective function $\Phi$ is regularized using a quadratic
penalty $\|\theta-\theta^k\|^2$.
The EM algorithm  can be viewed as  a generalized proximal point algorithm
where the quadratic regularization is replaced by a  Kullback-type
penalty.
Note that this 
 presentation includes the interpretation of EM as an 
alternating optimisation algorithm
(Neal and Hinton 1998, Hathaway 1986 in the 
mixture context). 
Equation (\ref{proxit}) becomes 
\begin{equation} \label{kpem}
\theta^{k+1}=\arg\max\limits_{\theta\in \Theta^\prime} \left\{
L(\theta\mid {\bf y})- D(\theta,\theta^k| {\bf y})\right\},
\end{equation}
where $L(\theta\mid {\bf y})$ is the observed log-likelihood of 
Section \ref{emsec}. 
The penalty term $D(\theta,\theta^k| {\bf y})$ is the 
 Kullback-Leibler divergence $I$ between the two conditional densities
${\bf t}({\bf x}| {\bf y},\theta)$ and ${\bf t}({\bf x}| {\bf y},\theta^k)$
 as defined in (\ref{tc}),
\begin{equation}
\label{Ddist}
D(\theta, \theta' | {\bf y})=
I({\bf t}({\bf x}| {\bf y},\theta'),{\bf t}({\bf x}| {\bf y},\theta)) 
= I\!\!E\bigl[
\log \frac{{\bf t}({\bf x}| {\bf y},\theta')}{{\bf t}({\bf x}|
{\bf y},\theta)}| {\bf y};\theta'\bigr].
\end{equation}

This quantity is well defined under unrestrictive regularity assumptions of
 the  parametrized conditional
 densities
${\bf t}({\bf x}| {\bf y},\theta)$ with $\theta \in \Theta$ (see
Celeux, Chr\'etien, Forbes and Mkhadri 1999 
for details).  
A question of importance is whether or not the following property holds,
\begin{eqnarray} \label{distlike}
D(\theta,\theta' | {\bf y})=0 &\Rightarrow & \theta' = \theta \;.
\end{eqnarray}
 Since the Kullback-Leibler divergence  is
strictly convex, nonnegative and is zero between identical distributions,
$D$ vanishes {\it iff} $t(\theta')= t(\theta)$. However, the operator
defined by $t(.)$ is not injective on the whole parameter space. Therefore,
(\ref{distlike}) does not generally hold  and 
the Kullback information does not {\em a priori} behave like a distance in
all directions of the parameter space. Howevere, in the mixture case,
 (\ref{distlike}) holds when the constraint 
$\Sum{\ell=1}{J} p_\ell =1$ is satisfied.
In addition, we proved in Celeux and al. (1999) that 
$t(.)$
is coordinate-wise injective which allows the Kullback measure to enjoy
some distance-like properties at least on coordinate subspaces.
More specifically, we proved (see Lemma 1 in Celeux and al. 1999) that for
any $\nu$ in $\{1,\ldots,J\}$ the operator $t(\theta_1,\ldots,\theta_{\nu-
1},
.,\theta_{\nu+1},\ldots, \theta_J)$ is injective.
The result below follows easily.
\begin{lemma}
\label{distprop}
The distance-like function $D(\theta,\theta^{\prime} \mid {\bf y})$ satisfies
the following properties
 
(i) $D(\theta,\theta^{\prime} \mid {\bf y})\geq 0$ for all $\theta^{\prime}$ a
nd $\theta$
in $\Theta$,
 
(ii) if $\theta$ and $\theta^\prime$ only differ in one coordinate,
$D(\theta,\theta^{\prime} \mid {\bf y})=0$ implies $\theta^{\prime}=\theta$.
\end{lemma}
This result is essential in proving convergence
properties (see Subsection \ref{cvsec})  of the CEMM algorithm. 


\vspace{.5cm}

The main difficulty 
in passing to a component-wise approach is the treatment
of the constraint 
\begin{equation}
\sum_{\ell=1}^{J} p_\ell=1. \nonumber
\end{equation}
Usually, a reduced parameter space is introduced, 
\begin{equation}
\Omega=\Big\{ 
\Big(p_1,\ldots,p_{J-1},\mu_1,\ldots,\mu_J,\Sigma_1,\ldots,\Sigma_J\Big) \Big\},
\end{equation}
the remaining proportion being deduced from the equality 
\begin{equation}
p_J=1-\Sum{\ell=1}{J-1} p_\ell, \nonumber
\end{equation}
see Redner and Walker (1984) for instance. This is obviously not
satisfactory in the context of coordinate-wise methods. 
A Lagrangian approach (ref ??)  seems more appropriate.
It
first provides an alternative interpretation
of the EM algorithm for mixtures, where the parameter $n$ is nothing but
the Lagrange multiplier associated to the proportion constraint.
The EM algorithm for mixtures is equivalent to the following 
 iteration,
\begin{equation}
\label{nouv}
\theta^{k+1}=\arg\max\limits_{\theta \in \Theta}
\left\{Q(\theta|\theta^k)-n \Big(\Sum{\ell=1}{J} p_\ell-1\Big)\right\}.
\end{equation}
Then, 
 Looking at the maximization steps (\ref{mcem}) and
(\ref{mem}) and using formulation (\ref{nouv}) for EM, we can easily
deduce the proximal representation of CEMM.
 We refer to Celeux and al. (1999) for a proof. 
\begin{proposition}
\label{proxrep}
The CEMM recursion is equivalent to a coordinate-wise generalized proximal
point
procedure of the type
\begin{equation}
\label{utu}
\theta^{k+1}=\arg\max\limits_{\theta \in \Theta_{k}}\left\{
L(\theta\mid {\bf y}) - n(\Sum{\ell=1}{J} p_\ell -1)
-D\big(\theta,\theta^k \mid y\big)\right\},
\end{equation}
where $\Theta_k$ is the  parameter set of the form
\begin{equation*}
\Theta_k=\Big\{ \theta \in \Theta \mid
\theta_\ell=\theta^k_\ell,\:\ell \neq j
\Big\}
\end{equation*}
with $j=k-\frac{k}{J}\rfloor J +1$.
\end{proposition}
We now establish a series of results concerning the CEMM iterations. 

\subsection{Convergence of CEMM}
\label{cvsec}
\begin{assumption}
\label{ass}
Let $\theta$ be any point in $\Theta$. Then, the level set 
\begin{equation}
\mathcal L_\theta=\Big\{ \theta^\prime \mid L(\theta^\prime | {\mathbf y}) \geq L(\theta | {\mathbf y}) \Big\} \nonumber
\end{equation}
is compact.
\end{assumption}
Let $\Lambda(\theta\mid {\bf y})$ be the modified log-likelihood function given by 
\begin{equation}
\Lambda(\theta\mid {\bf y}) = L(\theta\mid {\bf y}) -n (\Sum{\ell=1}{J} p_\ell-1). \nonumber
\end{equation}
This function first arised in the Lagrangian framework of Section
 \ref{proxsec}. It appears in the 
right-hand side of 
 equation (\ref{nouv}) when the Kullback-type penalty is omitted. 
\begin{proposition}
\label{nondec}
The sequence $\{\Lambda(\theta^k\mid {\bf y})\}_{k \in \mathbb N }$ is monotone non-decreasing, 
and satisfies
\begin{equation}
\label{inc}
\Lambda(\theta^{k+1}\mid {\bf y})-\Lambda(\theta^k\mid {\bf y})
\geq D(\theta^{k+1},\theta^k \mid {\bf y}).
\end{equation} 
\end{proposition}
{\bf Proof}.
>From iteration (\ref{utu}), we have
\begin{equation*}
\Lambda(\theta^{k+1}\mid {\bf y})-\Lambda(\theta^k\mid {\bf y})
\geq D(\theta^{k+1},\theta^k \mid {\bf y})-
D(\theta^k,\theta^k \mid {\bf y}).
\end{equation*}
The proposition follows from 
$D(\theta^{k+1},\theta^k \mid {\bf y})\geq 0$ and 
$D(\theta^k,\theta^k \mid {\bf y})= 0$.  
\hfill$\Box$

\begin{lemma}
\label{bound}
The sequence $\big\{\theta^k\big\}_{k\in \mathbb N}$ is bounded and  satisfies
\begin{equation}
\lim_{k \rightarrow \infty}\sum_{j=1}^J p_j^k=1. \label{deux}
\end{equation}
If in addition, $\{\Lambda(\theta^k | {\mathbf y})\}_{k\in \mathbb N}$ is 
bounded from above,  
\begin{equation} 
\lim_{k\rightarrow\infty}\|\theta^{k+1}-\theta^k\|=0 \label{trois}\;.
\end{equation}
\end{lemma}
{\bf Sketch of proof}. The fact that $\big\{\theta^k\big\}_{k\in \mathbb N}$
is bounded is straightforward from Proposition \ref{nondec} and Assumption
\ref{ass}. 
Equations $(\ref{deux})$ and (\ref{trois}) can be shown using 
Lemma \ref{distprop}
 and standard arguments on sequences (see Celeux and al. 1999 for
 details). \hfill$\Box$

The proof of the following theorem is in Appendix \ref{th1app}.
\begin{theorem}
Every accumulation point $\theta^*$ of the sequence 
$\big\{\theta^k\big\}_{k \in \mathbb N}$ satisfies one of the following two properties
\begin{itemize}
\item $\Lambda(\theta^*\mid {\bf y})=+\infty$ 
\item $\theta^*$ is 
a stationary point of the modified
log-likelihood function $\Lambda(\theta\mid {\bf y})$. 
\end{itemize}
\end{theorem}
The following result is direct consequence of Corollary 4.5 in Chr\'etien 
and Hero (1999).
\begin{corollary}
Assume that the modified log-likelihood function $\Lambda(\theta\mid {\bf y})$ is 
strictly concave in an open neighborhood of a stationary point 
of $\big\{\theta^k\big\}_{k \in \mathbb N}$.
 Then, the sequence $\big\{\theta^k\big\}_{k \in \mathbb N}$
 converges and its limit is a local maximizer of $\Lambda(\theta\mid {\bf y})$.
\end{corollary}

The main convergence result for the CEMM procedure is stated
below and its proof is given in Appendix \ref{th2app}. 
\begin{theorem}
Every accumulation point of the sequence 
$\big\{\theta^k\big\}_{k \in \mathbb N}$ is 
a stationary point of the log-likelihood function $L(\theta\mid {\mathbf y})$
on the set defined by the constraint $\sum_{\ell=1}^Jp_\ell=1$. 
\end{theorem}

\section{Numerical experiments}
\label{numsec}

The behaviors of EM, SAGE (as described in the Appendix of Celeux and al.
1999)
 and CEMM are compared  on the basis
of simulation experiments on univariate Gaussian mixtures with $J=3$
components.
First, we consider a mixture of well separated components with equal mixing
proportions $p_1=p_2=p_3=1/3$, means $\mu_1=0,\mu_2=3,\mu_3=6$ and
equal 
variances $\sigma_1^2=\sigma_2^2=\sigma_3^2=1$. We will  
refer to this mixture as the {\it well-separated} mixture.
Secondly, we consider a mixture of overlapping components 
with equal mixing
proportions $p_1=p_2=p_3=1/3$, means $\mu_1=0,\mu_2=3,\mu_3=3$ and
variances $\sigma_1^2=\sigma_2^2=1, \sigma^2_3=4$. 
This mixture will be referred to  as the {\it overlapping} mixture.

For the {\it well-separated}  mixture we consider a unique sample of size 
$n=300$ and perform the EM, SAGE and CEMM algorithms from
the following initial position:\[
p_1^0=p_2^0=p_3^0=1/3,\ \mu^0_1=\bar x - s, \mu^0_2= \bar x, \mu^0_3= \bar x +s, \sigma_1^0=\sigma_2^0=\sigma_3^0=s^2\]
where $\bar x$  and $s^2$ are respectively the empirical sample mean 
 and  variance.
Starting from this 
rather  favorable initial position, close to
 the true parameter values, the 
three algorithms converge to the same solution below 
\begin{eqnarray*}
\hat{p}_1&=&0.36, \hat{\mu}_1=0.00, \hat{\sigma}_1^2=1.10\\
\hat{p}_2&=&0.29, \hat{\mu}_2=2.96, \hat{\sigma}_2^2=0.38\\
\hat{p}_3&=&0.35, \hat{\mu}_3=5.90, \hat{\sigma}_3^2=1.10
\end{eqnarray*}
The performances of EM, SAGE and CEMM, in terms of speed,
 are compared on the basis of the {\it cycles} number
needed to reach the stationary value of the constraint log-likelihood.
\begin{center}
-------------------\\
Figure 1 about here\\
-------------------\\
\end{center}
A cycle corresponds to   the updating of all mixture components.
For EM, it consists of 
 a E-step (\ref{eem}) and a M-step (\ref{mem}). For SAGE,
it is the (J+1) iterations
 described in the Appendix. For CEMM, it consists of 
$J$ iterations described in (\ref{ecem}) and (\ref{mcem}). In each case,
a cycle of iterations requires the same number of
algebraic operations ,namely, $J$ updatings of the mixing proportions,
means and variance matrices and $J \times n$ updatings of the conditional
probabilities $t_{ij}(\theta)$.


Figure \ref{f1} displays the log-likelihood versus cycle for
EM, SAGE and CEMM in the  
{\it well-separated} mixture case. As expected, when  starting
from a good initial position in a well separated mixture situation,
EM converges  rapidly to a local  maximum of the  likelihood. 
Moreover, EM
outperforms SAGE and CEMM in this example.

For the {\it overlapping} mixture, we consider two different samples of size
$n=300$ and performed the EM, SAGE and CEMM algorithms from
the following initial position:
\[p_1^0=p_2^0=p_3^0=1/3,\ \mu^0_1=0.0, \mu^0_2= 0.1, \mu^0_3= 0.2, \sigma_1^0=\sigma_2^0=\sigma_3^0=1.0,\]
which is 
far from the true parameter values. For 
the first sample, the three algorithms converge to the same solution 
\begin{eqnarray*}
\hat{p}_1&=&0.65, \hat{\mu}_1=0.85, \hat{\sigma}_1^2=1.28\\
\hat{p}_2&=&0.19, \hat{\mu}_2=3.32, \hat{\sigma}_2^2=0.26\\
\hat{p}_3&=&0.16, \hat{\mu}_3=5.67, \hat{\sigma}_3^2=2.10.
\end{eqnarray*}
\begin{center}
--------------------------\\
Figures 2 and 3 about here\\
--------------------------\\
\end{center}
Figure \ref{f3} displays the log-likelihood versus cycle for
EM, SAGE and CEMM for the  first
sample of the {\it overlapping}  mixture. In this situation, EM appears to
converge slowly so that  SAGE and especially CEMM show a significant 
improvement of convergence speed.

For the second sample, starting from the same position,
SAGE and CEMM both  converge to the solution below 
\begin{eqnarray*}
\hat{p}_1&=&0.61, \hat{\mu}_1=0.85, \hat{\sigma}_1^2=1.62\\
\hat{p}_2&=&0.13, \hat{\mu}_2=3.00, \hat{\sigma}_2^2=0.52\\
\hat{p}_3&=&0.26, \hat{\mu}_3=4.27, \hat{\sigma}_3^2=4.29,
\end{eqnarray*}
while EM proposes the following solution, after 1000 cycles,
\begin{eqnarray*}
\hat{p}_1&=&0.61, \hat{\mu}_1=0.83, \hat{\sigma}_1^2=1.60\\
\hat{p}_2&=&0.16, \hat{\mu}_2=2.98, \hat{\sigma}_2^2=0.62\\
\hat{p}_3&=&0.22, \hat{\mu}_3=4.58, \hat{\sigma}_3^2=4.29.
\end{eqnarray*}
Figure \ref{f4} displays the log-likelihood versus cycle for
EM, SAGE and CEMM for the second
sample of the {\it overlapping} mixture. The same conclusions hold for
this sample. The 
CEMM algorithm is the faster while  EM is really slow, the 
correspondant local maximum of the likelihood not being reached after 1000
ierations. 

Moreover,  it appears that the implemented version of the SAGE algorithm
is less beneficial than CEMM for situations in which EM converges slowly.
A possible reason for this behavior of SAGE 
is that the $(J+1)$th iteration of SAGE involves the whole complete
data structure, 
whereas CEMM iterations never
need the whole complete data space
as missing data space. 

\section{Concluding remarks}
\label{discsec}

We presented a component-wise
EM algorithm for finite identifiable mixtures of distributions (CEMM) and
proved  convergence properties similar to that of standard EM.
As illustrated in section \ref{numsec}, 
numerical experiments suggest that CEMM and EM have complementary 
performances. The CEMM algorithm
  is of poor interest when EM  convergence is fast but
shows significant improvement when EM encounters slow convergence
rate. Thus, CEMM may be useful in many contexts. 
An intuitive explanation of our procedure performances is that the 
component-wise strategy prevents the algorithm  from staying too long at 
critical points (typically saddle points)
 where standard EM is likely to get trapped. More theoretical
investigations would be interesting but are beyond the scope of the 
present  paper. 

Other futur directions of research include   the use of relaxation, as in 
Chr\'etien  and Hero (1998b), for accelerating
  CEMM, and the possibility of using 
varying/adaptative orders to update the components.  

\newpage
\appendix

\section{Proof of Theorem 1}
\label{th1app}

\section{Proof of Theorem 2}
 \label{th2app}
Let $\theta^*$ be an accumulation point of $\{\theta^k\}_{k\in \mathbb N}$.
Note that $\theta^*$ lies in
$\Theta^\prime=\Big\{\theta\in \Theta \mid \sum_{\ell=1}^Jp_\ell=1 \Big\}. $
Take any vector $\delta$ such that
$\theta^*+\delta$ lies in $\Theta^\prime$. Since $\Theta^\prime$ is affine, an
y point $\theta_t=
\theta^*+t\delta$, $t\in \mathbb R$ also lies in $\Theta^\prime$. The
directional derivative of $\Lambda$ at $\theta^*$ in the direction $\delta$ is
 obviously null.
It is given by
\begin{equation}
(0=)\Lambda^\prime(\theta^*;\delta \mid {\mathbf y})=\lim_{t\rightarrow 0^+}
\frac{\Lambda(\theta^*\mid {\mathbf y})-\Lambda(\theta^*+t\delta \mid {\mathbf
 y})}{t}, \nonumber
\end{equation}
which is equal to
\begin{equation}
\Lambda^\prime(\theta^*;\delta \mid {\mathbf y})=\lim_{t\rightarrow 0^+}
\frac{L(\theta^*\mid {\mathbf y})-L(\theta^*+t\delta \mid {\mathbf y})+
c(\theta^*) -c(\theta^*+t\delta)}{t}, \nonumber
\end{equation}
where $c(\theta)=n\Big(\sum_{\ell=1}^Jp_\ell-1 \Big)$. Since $\theta^*+t\delta
$ lies
in $\Theta^\prime$ for all nonnegative $t$, $c(\theta^*+t\delta)=c(\theta^*)=0
$, and
we obtain
\begin{equation}
\Lambda^\prime(\theta^*;\delta \mid {\mathbf y})=L^\prime(\theta^*;\delta \mid
 {\mathbf y}).  \nonumber
\end{equation}
Thus,
\begin{equation}
L^\prime(\theta^*;\delta \mid {\mathbf y})=0
\end{equation}\\\\

{\bf References}
\begin{description}
\item[   ] Celeux, G., Chr\'etien, S., Forbes, F.,
 and Mkhadri, A. (1999), "A Component-Wise {EM} Algorithm for Mixtures",
{\em Technical Report}, Inria 3746.
(http://www.inria.fr/RRRT/publications-fra.html).
\item[   ] Chr\'etien, S., and Hero, A. O.(1998a), "Acceleration of the {EM} algorithm
via proximal point iterations", {\em IEEE International Symposium on Information Theory}, MIT Boston.
\item[   ] Chr\'etien, S. and Hero, A. O. (1998b), "Generalized proximal point algorithms and bundle implementation", {\em Technical Report}, CSPL 313, The University of Michigan, Ann Arbor, USA.
\item[   ] Chr\'etien, S., and Hero, A. O. (1999), "Kullback proximal algorithms for maximum likelihood estimation", {\em Technical Report}, Inria 3756.
(http://www.inria.fr/RRRT/publications-fra.html).
\item[   ] Ciarlet, P. G. (1988), "{\em Introduction to numerical linear algebra and optimization}", Cambridge Texts in Applied Mathematics: Cambridge University Press.
\item[   ] Dempster, A. P., Laird, N. M., and Rubin, D. B. (1977), "Maximum likelihood for incomplete data via the {EM} algorithm (with discussion)",{\em J. Roy. Stat. Soc. Ser. B}, 39, 1-38.
\item[   ] Fessler, J. A., and Hero, A. O.(1994), "Space-Alternating generalized expectation-maximisation algorithm", {\em IEEE Trans. Signal Processing}, 42, 2664-2677.
\item[   ] Fessler, J. A., and Hero, A. O. (1995), "Penalized maximum-likelihood image reconstruction using space-Alternating generalized {EM} algorithms", {\em IEEE Trans. Image Processing}, 4, 1417-1429.
\item[   ] Hathaway, R.J. (1986), "Another interpretation of {EM} algorithm for mixture distributions", {\em Statist. and Probab. Letters}, 4, 53-56.
\item[   ] Hero, A. O., and Fessler, J. A. (1995), "Convergence in norm for {EM}-type algorithms", {\em Statistica Sinica}, 5, 41-54.
\item[   ] Ibragimov, I. A., and Khas'minskij, R. Z. (1981), "{\em Asymptotic theory of estimation}", Springer-Verlag.
\item[   ] Jamshidian, M. and Jennrich, R. I.", "Conjugate gradient acceleration of the {EM} algorithm", {\em J. Amer. Stat. Ass.}, 88, 221-228.
\item[   ] Kaufman, L. (1987), "Implementing and accelerating the {EM} algorithm for positron emission tomography", {\em IEEE Trans. Medical Image}, 6, 37-51.
\item[   ] Lewitt, R. M., and Muehllehner (1986), "Accelerated iterative reconstruction for positron emission tomography based on the {EM} algorithm for maximum likelihood estimation",{\em IEEE Trans. on Medical Image}, 5, 16-22.
\item[   ] Liu, C., and Rubin, D.B. (1994), The {ECME} algorithm: A simple extension of {EM} and {ECM} with faster monotone convergence", {\em Biometrika}, 81, 633-648.
\item[   ] Liu, C., and Sun, D. X. (1997), "Acceleration of {EM} algorithm for mixtures models using {ECME}", {\em ASA Proceedings of The Stat. Comp. Session}, 109-114.
\item[   ] Liu, C., Rubin, D.B., and Wu, Y. (1998), "Parameter expansion to accelerate {EM}: the {PX-EM} algorithm", {\em Biometrika}, 755-770. 
\item[   ] Louis, T. A. (1982), "Finding the observed information matrix when using the {EM} algorithm", {\em J. Roy. Stat. Soc. Ser. B}, 44, 226-233.
\item[   ] McLachlan, G. J., and Krishnam, T.(1997), "{\em The {EM} algorithm and extensions}", New York-London. Sydney-Toronto: John Wiley and Sons, Inc.
\item[   ] Meilijson, I. (1989), "A fast improvement to the {EM} algorithm in its own terms",{\em J. Roy. Stat. Soc. Ser. B}, 51, 127-138.
\item[   ] Meng, X.-L., and Rubin, D.B., (1993). Maximum likelihood estimation via the ECM algorithm: A general framework, {\em Biometrika}, 80, 267-278.
\item[   ] Meng, X.-L., and van Dyk, D. A. (1997), "The {EM} algorithm - an old folk song sung to a fast new tune (with discussion)", {\em J. Roy. Stat. Soc. Ser. B}, 59, 511-567.
\item[   ] Meng, X.-L., and van Dyk, D. A. (1998), "Fast {EM}-type implementations for mixed effects models",{\em J. Roy. Stat. Soc. Ser. B}, 60, 559-578.
\item[   ] Neal, R. N., and Hinton, G. E. (1998), "{\em A view of the {EM} algorithm that justifies incremental, sparse and other variants}", Learning in Graphical Models, Jordan, M.I. (Editor), Dordrencht, Kluwer Academic Publishers.
\item[   ] Pilla, R. S., and Lindsay, B. G. (1996),"Faster {EM} methods in high-dimensional finite mixtures", In "{Proceedinds of the Statistical Computing Section}, 166-171, Alexandria, Virginia. ASA.
\item[   ] Redner, R. A., and Walker, H. F. (1984), "Mixture densities, maximum likelihood and the {EM} algorithm",{\em SIAM Review}, 26, 195-239.
\item[   ] Rockafellar, R. T. (1976a), "Augmented Lagrangians and application of the proximal point algorithm in convex programming", {\em Mathematics of Operations Research}, 17, 96-116.
\item[   ] Rockafellar, R. T. (1976b), "Monotone operators and the proximal point algorithm", {\em SIAM Journal on  Control and Optimization}, 14, 877-898.
\item[   ] Teboulle, M.(1992), "Entropic proximal mappings with application to nonlinear programming", {\em Mathematics of Operations Research}, 17, 670-690.
\item[   ] Teboulle, M. (1997), "Convergence of proximal-like algorithms", {\em SIAM Journal on  Control and Optimization}, 7, 1069-1083.
\end{description}
  
\newpage

\begin{figure}
\begin{center}
\makebox[8 cm][l]{
\vbox to 7 cm{
\vfill
\includegraphics{fig1.ps}
}}
\caption{Comparison of log-likelihood versus cycle for
EM (full line), SAGE (dashed line) and CEMM (dotted line)
in  the {\it well-separated}  mixture case.}
\label{f1}
\end{center}
\end{figure}

\begin{figure}
\begin{center}
\makebox[8 cm][l]{
\vbox to 7 cm{
\vfill
\includegraphics{fig3.ps}
}}
\caption{Comparison of log-likelihood versus cycle for
EM (full line), SAGE (dashed line) and CEMM (dotted line)
in the {\it overlapping} mixture case (first sample).}
\label{f3}
\end{center}
\end{figure}

\begin{figure}
\begin{center}
\makebox[8 cm][l]{
\vbox to 7 cm{
\vfill
\includegraphics{fig4.ps}
}}
\caption{Comparison of log-likelihood versus cycle for
EM (full line), SAGE (dashed line) and CEMM (dotted line)
in the {\it overlapping} mixture case (second sample).}
\label{f4}
\end{center}
\end{figure}
\end{document}